\documentclass[twocolumn,aps,pra,amsmath,amssymb,notitlepage]{revtex4-1}
\usepackage[latin9]{inputenc}
\usepackage{mathrsfs}
\usepackage{amsmath}
\usepackage{amssymb}
\usepackage{graphicx}
\usepackage{esint}

\makeatletter

\newcommand*\LyXThinSpace{\,\hspace{0pt}}

\usepackage{epsfig}
\usepackage{bm}
\usepackage{dcolumn}
\usepackage{color}

\makeatother

\begin{document}
\title{Thermal destabilization of self-bound ultradilute quantum droplets}
\author{Jia Wang, Hui Hu, and Xia-Ji Liu}
\affiliation{Centre for Quantum Technology Theory, Swinburne University of Technology,
Melbourne, Victoria 3122, Australia}
\date{\today}
\begin{abstract}
We theoretically investigate the temperature effect in a Bose-Bose
mixture with attractive inter-species interactions, in the regime
where a self-bound ultradilute quantum droplet forms due to the subtle
balance between the attractive mean-field force and the repulsive
force provided by Lee-Huang-Yang quantum fluctuations. We find that
in contrast to quantum fluctuations, thermal fluctuations destabilize
the droplet state and completely destroy it above a threshold temperature.
We show that the threshold temperature is determined by the intra-species
interaction energy. For a three-dimensional Bose-Bose mixture, the
threshold temperature is less than one-tenth of the Bose-Einstein
condensation temperature under the typical experimental conditions.
With increasing temperature, the droplet's equilibrium density gradually
decreases and can be reduced by several tens of percent upon reaching
the threshold temperature. We also consider a one-dimensional quantum
droplet and find a similar destabilization effect due to thermal fluctuations.
The threshold temperature in one dimension is roughly set by the binding
energy of the inter-species dimer. The pronounced thermal instability
of a self-bound quantum droplet predicted in our work could be examined
in future experiments, by measuring the temperature dependence of
its central density and observing its sudden disappearance at the
threshold temperature.
\end{abstract}
\maketitle

\section{Introduction}

In his classical textbook ``\textit{The Universe in a Helium Droplet}''
\citep{Volovik2009}, Volovik described an interesting autonomously
isolated quantum system of helium nanodroplet, without any interaction
with the surrounding environment. It can be in equilibrium at \emph{zero}
external pressure (i.e., $P=0$) in empty space, with an equilibrium
particle density $n_{\textrm{eq}}\sim2\times10^{22}$cm$^{-3}$ \citep{Harms1998,Barranco2006,Gessner2019}
determined by the balance between the attractive interatomic interaction
($\propto n$) and the repulsive zero-point quantum motions of helium
atoms ($\propto n^{3/2}$). As a \emph{negative} chemical potential
$\mu<0$ is needed to prevent self-evaporation, the droplet formation
is generally impossible in the well-studied single-component weakly
interacting Bose gas of utracold alkali-metal atoms \citep{Dalfovo1999},
where the attractive short-range inter-particle interactions lead
to a mean-field collapse \citep{Donley2001}.

The situation, however, can dramatically change if one considers a
binary Bose mixture with both short-range attractive inter-species
interactions ($g_{12}<0$) and repulsive intra-species interactions
($g>0$), as suggested by Petrov in his seminal work \citep{Petrov2015}.
When $g_{12}<-g$, although the mean-field theory predicts a collapsing
state, the inclusion of Lee-Huang-Yang (LHY) quantum fluctuations
\citep{LeeHuangYang1957} (i.e., the zero-point oscillations) turns
out to stabilize the mechanical collapse and leads to the formation
of an ultradilute droplet at equilibrium density $n_{\textrm{eq}}\sim10^{14}$cm$^{-3}$,
which is about eight orders of magnitude more dilute than a liquid
$^{4}$He droplet. Petrov's ground-breaking idea has now been successfully
confirmed experimentally with homonuclear $^{39}$K-$^{39}$K mixtures
\citep{Cabrera2018,Cheiney2018,Semeghini2018,Ferioli2019} and heteronuclear
$^{41}$K-$^{87}$Rb mixtures \citep{DErrico2019}. Ultradilute quantum
droplets have also been observed in scalar weakly-interacting Bose
gases with anisotropic long-range dipolar interactions \citep{FerrierBarbut2016,Schmitt2016,Chomaz2016,Bottcher2019},
following more closely the picture by Volovik. These rapid experimental
developments open a new fast-moving research direction, where several
intriguing many-body effects can be predicted theoretically \citep{Petrov2016,Baillie2016,Wachtler2016,Li2017,Cappellaro2018,Chiquillo2018,Astrakharchik2018,Cui2018,Staudinger2018,Ancilotto2018,Parisi2019,Aybar2019,Cikojevic2019,Chiquillo2019,Tylutki2020,Cikojevic2020,Wang2020}
and examined experimentally \citep{Cabrera2018,Cheiney2018,Semeghini2018,Ferioli2019,DErrico2019,FerrierBarbut2016,Schmitt2016,Chomaz2016,Bottcher2019},
beyond the existing paradigm of helium nanodroplets \citep{Harms1998,Barranco2006,Gessner2019}.

In this work, we would like to understand how quantum droplets' \emph{bulk}
properties are affected by a small but nonzero temperature, which
always exists experimentally. This issue is rarely addressed in the
past literature, presumably due to the lack of a useful microscopic
theory of quantum droplets from the first principle. Here, we overcome
such a difficulty by extending a most recently developed pairing description
of the droplet state \citep{Hu2020a,Hu2020b,Hu2020c} to finite temperatures.
For simplicity, we follow Petrov's binary Bose mixture model of quantum
droplets \citep{Petrov2015}, to avoid subtly treating the long-range
anisotropic dipolar interactions in dipolar droplets.

We find that thermal fluctuations generally destabilize the droplet.
This tendency is natural to understand. As temperature increases,
the atomic motion becomes increasingly significant, and the droplet
may fail to maintain its zero-pressure state. We observe that the
threshold temperature for the complete destruction of the quantum
droplet is typically set by the intra-species interaction energy.
For a weakly interacting Bose-Bose mixture in three dimensions, the
interaction energy is small, so the threshold temperature could be
less than one-tenth of the condensation temperature under the current
experimental conditions \citep{Cabrera2018,Semeghini2018}. Remarkably,
despite this low threshold temperature, the equations of state of
quantum droplets still show a strong temperature dependence. In particular,
with increasing temperature, the droplet's equilibrium density can
be reduced by several tens of percent upon reaching the threshold
temperature. We also consider the droplet state in a Bose-Bose mixture
in one dimension and find similar thermal destabilization due to thermal
fluctuations.

The rest of the paper is organized as follows. In the next section
(Sec. II), we introduce the model Hamiltonian and present a microscopic
pairing theory of quantum droplets at low but finite temperature based
on the conventional Bogoliubov theory \citep{Larsen1963}. In Sec.
III, we discuss the thermal destabilization of quantum droplets in
three dimensions. The quantum depletion and thermal depletion are
calculated to validate the applicability of the Bogoliubov theory.
In Sec. IV, we consider one-dimensional quantum droplets and show
that the thermal destabilization effect is universal for the droplet
state in different dimensions. Finally, Sec. V is devoted to the conclusions
and outlooks.

\section{Pairing theory at finite temperature}

To highlight the essential temperature effect, we consider the \emph{simplest}
possible homonuclear Bose-Bose mixture, with equal repulsive intra-species
interactions (i.e., $g_{11}=g_{22}=g$) and attractive inter-species
interactions ($g_{12}$). In this case, we have equal population in
each species and equal chemical potential $\mu$. In real space, the
system is described by a grand canonical model Hamiltonian, $\hat{K}=\int d\mathbf{x}[\mathscr{H}+\mathscr{H}_{\textrm{intra}}+\mathscr{H}_{\textrm{inter}}]$,
with 
\begin{eqnarray}
\mathscr{H}_{0} & = & \sum_{i=1,2}\hat{\phi}_{i}^{\dagger}\left(\mathbf{x}\right)\left[-\frac{\hbar^{2}\nabla^{2}}{2m}-\mu\right]\hat{\phi}_{i}\left(\mathbf{x}\right),\\
\mathscr{H}_{\textrm{intra}} & = & \frac{g}{2}\sum_{i=1,2}\hat{\phi}_{i}^{\dagger}\left(\mathbf{x}\right)\hat{\phi}_{i}^{\dagger}\left(\mathbf{x}\right)\hat{\phi}_{i}\left(\mathbf{x}\right)\hat{\phi}_{i}\left(\mathbf{x}\right),\\
\mathscr{H}_{\textrm{inter}} & = & -\frac{\hat{\Delta}^{\dagger}\hat{\Delta}}{g_{12}}-\left[\hat{\Delta}\hat{\phi}_{1}^{\dagger}\left(\mathbf{x}\right)\hat{\phi}_{2}^{\dagger}\left(\mathbf{x}\right)+\textrm{H.c.}\right],
\end{eqnarray}
where in the last Hamiltonian density for inter-species interactions,
we have used the Hubbard-Stratonovich transformation to decouple the
four-fermion interaction term through the introduction of a pairing
field $\hat{\Delta}(\mathbf{x})$ \citep{Hu2020b,Hu2020c}. $\hat{\phi}_{i}(\mathbf{x})$
and $\hat{\phi}_{i}^{\dagger}(\mathbf{x})$ ($i=1,2$) are annihilation
and creation field operators for the $i$-species bosons with mass
$m_{1}=m_{2}=m$. 

In this work, we consider both three-dimensional and one-dimensional
binary mixtures. In three dimensions, the short-range \emph{contact}
inter-particle interactions used in the model Hamiltonian are unphysical
in the large-momentum and high-energy limit, as reflected by the well-known
ultraviolet divergence. The divergence can be removed by the standard
regularization procedure: we simply re-express the bare interaction
strengths $g_{ij}$ in terms of the $s$-wave scattering lengths $a_{ij}$,
i.e.,
\begin{equation}
\frac{1}{g_{ij}}=\frac{m}{4\pi\hbar^{2}a_{ij}}-\frac{1}{\mathcal{V}}\sum_{\mathbf{k}}\frac{m}{\hbar^{2}\mathbf{k}^{2}},\label{eq:Reg3d}
\end{equation}
where $\mathcal{V}$ is the volume of the system (or the length of
the system in the one-dimensional case). The interaction regularization
is not required in one dimension. There, the interaction strengths
$g_{ij}$ are related to the $s$-wave scattering length via 
\begin{equation}
g_{ij}=-\frac{2\hbar^{2}}{ma_{ij}}.
\end{equation}

\subsection{Bogoliubov theory with pairing}

The pairing theory of quantum droplets in a binary Bose mixture has
been discussed in detail in the previous works \citep{Hu2020a,Hu2020b,Hu2020c}.
Here, for self-containedness we briefly review the theory and extend
it to the finite temperature case. In the weakly interacting regime,
we take the Bogoliubov approximation to rewrite the bosonic field
operators \citep{Larsen1963,Fetter1972,Griffin1996},
\begin{equation}
\hat{\phi}_{i}\left(\mathbf{x}\right)=\phi_{c}\left(\mathbf{x}\right)+\delta\hat{\phi}_{i}\left(\mathbf{x}\right),
\end{equation}
where $\delta\hat{\phi}_{i}$ is considered as small fluctuation around
the condensate wave-function $\phi_{c}(\mathbf{x})$. At the same
level of approximation, we also take a a static $c$-number function
for the pairing field \citep{Hu2020a,Hu2020b,Hu2020c}, i.e., 
\begin{equation}
\hat{\Delta}\left(\mathbf{x}\right)=\Delta\left(\mathbf{x}\right),
\end{equation}
and determine it \emph{variationally}. As we focus on the ground state,
both the condensate wave-function $\phi_{c}(\mathbf{x})$ and the
pairing function $\Delta(\mathbf{x})$ can be chosen as real and non-negative
functions. By expanding the model Hamiltonian in terms of $\delta\hat{\phi}_{i}^{\dagger}$
and $\delta\hat{\phi}_{i}$ and truncate it to the second order, we
find that a quadratic form,
\begin{eqnarray}
\hat{K}_{B} & = & \sum_{i=1,2}\int d\mathbf{x}\left[\delta\hat{\phi}_{i}^{\dagger}\mathscr{L}\delta\hat{\phi}_{i}+\left(\frac{C}{2}\delta\hat{\phi}_{i}^{\dagger}\delta\hat{\phi}_{i}^{\dagger}+\textrm{H.c.}\right)\right]\nonumber \\
 &  & -\int d\mathbf{x}\left[\left(\Delta\delta\hat{\phi}_{1}^{\dagger}\delta\hat{\phi}_{2}^{\dagger}+\textrm{H.c.}\right)+\frac{C^{2}}{g}+\frac{\Delta^{2}}{g_{12}}\right],\label{eq:BogHami}
\end{eqnarray}
provided that $\phi_{c}(\mathbf{x})$ and $\Delta(\mathbf{x})$ satisfy
a Gross-Pitaevskii (GP) equation,
\begin{equation}
\left[-\frac{\hbar^{2}\nabla^{2}}{2m}-\mu+C\left(\mathbf{x}\right)-\Delta\left(\mathbf{x}\right)\right]\phi_{c}\left(x\right)=0,\label{eq:GPE}
\end{equation}
where we have defined the short-hand notations,
\begin{eqnarray}
C\left(\mathbf{x}\right) & \equiv & g\phi_{c}^{2}\left(\mathbf{x}\right),\\
\mathscr{L} & \equiv- & \frac{\hbar^{2}\nabla^{2}}{2m}-\mu+2C\left(\mathbf{x}\right).
\end{eqnarray}
The function $C(\mathbf{x})$ can be simply viewed as the intra-species
interaction energy density. The quadratic form of the model Hamiltonian
is straightforward to diagonalize, by a linear real-space Bogoliubov
transformation \citep{Fetter1972,Griffin1996},
\begin{eqnarray}
\delta\hat{\phi}_{i}\left(\mathbf{x}\right) & = & \sum_{n}\left[u_{ni}\left(\mathbf{x}\right)\hat{\alpha}_{n}+v_{ni}^{*}\left(\mathbf{x}\right)\hat{\alpha}_{n}^{\dagger}\right],\\
\delta\hat{\phi}_{i}^{\dagger}\left(\mathbf{x}\right) & = & \sum_{n}\left[u_{ni}^{*}\left(\mathbf{x}\right)\hat{\alpha}_{n}^{\dagger}+v_{ni}\left(\mathbf{x}\right)\hat{\alpha}_{n}\right],
\end{eqnarray}
where $\hat{\alpha}_{n}^{\dagger}$ and $\hat{\alpha}_{n}$ are creation
and annihilation field operators of Bogoliubov quasiparticles, and
$u_{ni}(\mathbf{x})$ and $v_{ni}(\mathbf{x})$ are the corresponding
quasiparticle wave-functions. The truncated Bogoliubov Hamiltonian
then becomes \citep{Hu2020c},
\begin{eqnarray}
\hat{K}_{B} & = & \varOmega_{0}+\sum_{n}E_{n}\hat{\alpha}_{n}^{\dagger}\hat{\alpha}_{n},\label{eq:BogHamiDiag}\\
\varOmega_{0} & = & -\int d\mathbf{x}\left[\frac{C^{2}}{g}+\frac{\Delta^{2}}{g_{12}}+\sum_{ni}E_{n}\left|v_{ni}\left(\mathbf{x}\right)\right|^{2}\right],\label{eq:OmegaT0}
\end{eqnarray}
if $u_{ni}(\mathbf{x})$ and $v_{ni}(\mathbf{x})$ obey the Bogoliubov
equations,
\begin{eqnarray}
\mathscr{L}u_{ni}+C\left(\mathbf{x}\right)v_{ni}-\Delta\left(\mathbf{x}\right)v_{n,3-i} & = & +E_{n}u_{ni},\label{eq:BogoliubovEQ1}\\
\mathscr{L}^{*}v_{ni}+C\left(\mathbf{x}\right)u_{ni}-\Delta^{*}\left(\mathbf{x}\right)u_{n,3-i} & = & -E_{n}v_{ni},\label{eq:BogoliubovEQ2}
\end{eqnarray}
where $E_{n}\geq0$ is the energy of Bogoliubov quasi-particles. It
is easy to check that the zero-mode with $E=0$ has the form $u_{1}=u_{2}=+\phi_{c}(\mathbf{x})$
and $v_{1}=v_{2}=-\phi_{c}(\mathbf{x})$, which is precisely the condensate
mode of the GP equation and hence should be excluded. From the diagonalized
Hamiltonian (\ref{eq:BogHamiDiag}), it is straightforward to write
down the thermodynamic potential at \emph{finite} temperature ($\beta=1/k_{B}T$),
\begin{equation}
\varOmega=\varOmega_{0}+\frac{1}{\beta}\sum_{n}\ln\left[1-\exp\left(-\beta E_{n}\right)\right],\label{eq:OmegaT1}
\end{equation}
from which, we determine the variational pairing function $\Delta(\mathbf{x})$
through the \emph{functional} minimization, i.e., 
\begin{equation}
\frac{\delta\varOmega\left[\mu,\Delta\left(\mathbf{x}\right)\right]}{\delta\Delta(\mathbf{x})}=0.
\end{equation}
Once $\Delta(\mathbf{x})$ is found, we calculate the total number
of atoms, $N=-\partial\varOmega/\partial\mu$, and consequently the
free energy $F=\Omega+\mu N$. In the thermodynamic limit, the free
energy per particle takes a global minimum as a function of the density
$n=N/\mathcal{V}$ when the system is in the self-bound droplet state,
i.e., $\partial(F/N)/\partial n=0$, which is equivalent to the zero-pressure
condition, 
\begin{equation}
P=-\frac{\varOmega}{\mathcal{V}}=\frac{n^{2}}{\mathcal{V}}\frac{\partial\left(F/N\right)}{\partial n}=0.
\end{equation}

It is worth noting that the above Bogoliubov theory is applicable
at low temperatures, where the depletion from the condensate, due
to either quantum fluctuations or thermal fluctuations, should be
sufficiently small. For this purpose, one may need to explicitly examine
the quantum depletion $N_{\textrm{qd}}$ and thermal depletion $N_{\textrm{th}}$.
These two quantities can be evaluated by taking the average of the
number operators ($i=1,2$), 
\begin{equation}
\left\langle \delta\hat{\phi}_{i}^{\dagger}\delta\hat{\phi}_{i}\right\rangle =\sum_{n}\left[\left|v_{ni}\right|^{2}+\left(\left|u_{ni}\right|^{2}+\left|v_{ni}\right|^{2}\right)\left\langle \hat{\alpha}_{n}^{\dagger}\hat{\alpha}_{n}\right\rangle \right],
\end{equation}
and we obtain,
\begin{eqnarray}
N_{\textrm{qd}} & = & \int d\mathbf{x}\sum_{ni}\left|v_{ni}\left(\mathbf{x}\right)\right|^{2},\\
N_{\textrm{th}} & = & \int d\mathbf{x}\sum_{ni}\left[\left|u_{ni}\left(\mathbf{x}\right)\right|^{2}+\left|v_{ni}\left(\mathbf{x}\right)\right|^{2}\right]f_{B}\left(E_{n}\right),
\end{eqnarray}
where $f_{B}(E)=1/(e^{\beta E}-1)$ is the Bose-Einstein distribution
function.

\subsection{Bulk properties of quantum droplets in the thermodynamic limit}

For simplicity, in this work we consider a sufficiently large droplet,
where the edge effect can be safely neglected. We therefore have \emph{constant}
condensate wave-function $\phi_{c}$, intra-species interaction energy
$C$, and pairing parameter $\Delta$. The GP equation (\ref{eq:GPE})
then leads to the relation $C=\mu+\Delta$. In this case, the quasi-particle
wave-functions, $u_{\mathbf{k}i}(\mathbf{x})=u_{\mathbf{k}i}e^{i\mathbf{k}\cdot\mathbf{x}}/\sqrt{\mathcal{V}}$
and $v_{\mathbf{k}i}(\mathbf{x})=v_{\mathbf{k}i}e^{i\mathbf{k}\cdot\mathbf{x}}/\sqrt{\mathcal{V}}$,
are plane waves with momentum $\mathbf{k}$ and energy $E_{\mathbf{k}}$.
The Bogoliubov equations (\ref{eq:BogoliubovEQ1}) and (\ref{eq:BogoliubovEQ2})
in momentum space take the form,
\begin{equation}
\left[\begin{array}{cccc}
B_{\mathbf{k}} & 0 & C & -\Delta\\
0 & B_{\mathbf{k}} & -\Delta & C\\
C & -\Delta & B_{\mathbf{k}} & 0\\
-\Delta & C & 0 & B_{\mathbf{k}}
\end{array}\right]\left[\begin{array}{c}
u_{\mathbf{k}1}\\
u_{\mathbf{k}2}\\
v_{\mathbf{k}1}\\
v_{\mathbf{k}2}
\end{array}\right]=E_{\mathbf{k}}\left[\begin{array}{c}
+u_{\mathbf{k}1}\\
+u_{\mathbf{k}2}\\
-v_{\mathbf{k}1}\\
-v_{\mathbf{k}2}
\end{array}\right],
\end{equation}
where $B_{\mathbf{k}}\equiv\varepsilon_{\mathbf{k}}+C+\Delta$ with
$\varepsilon_{\mathbf{k}}=\hbar^{2}\mathbf{k}^{2}/(2m)$. By defining
collectively $\mathbf{u}_{\mathbf{k}}\equiv[u_{\mathbf{k}1},u_{\mathbf{k}2}]^{T}$
and $\mathbf{v}_{\mathbf{k}}\equiv[v_{\mathbf{k}1},v_{\mathbf{k}2}]^{T}$,
it is easy to check that,
\begin{eqnarray}
\mathbf{u}_{\mathbf{k}} & = & -\frac{\mathbf{M}}{B_{\mathbf{k}}-E_{\mathbf{k}}}\mathbf{\mathbf{v}_{\mathbf{k}}},\\
\mathbf{v}_{\mathbf{k}} & = & -\frac{\mathbf{M}}{B_{\mathbf{k}}+E_{\mathbf{k}}}\mathbf{\mathbf{u}_{\mathbf{k}}},
\end{eqnarray}
where 
\begin{equation}
\mathbf{M}=\left[\begin{array}{cc}
C & -\Delta\\
-\Delta & C
\end{array}\right]
\end{equation}
is a 2 by 2 matrix. It is straightforward to show that the quasi-particle
wave-functions satisfy \citep{Hu2020c},
\begin{align}
u_{\mathbf{k}1}^{2} & =u_{\mathbf{k}2}^{2}=\frac{1}{4}\left(\frac{B_{\mathbf{k}}}{E_{\mathbf{k}}}+1\right),\\
v_{\mathbf{k}1}^{2} & =v_{\mathbf{k}2}^{2}=\frac{1}{4}\left(\frac{B_{\mathbf{k}}}{E_{\mathbf{k}}}-1\right),
\end{align}
where the dispersion relation $E_{\mathbf{k}}$ can take two branches,

\begin{eqnarray}
E_{\mathbf{k}-} & = & \sqrt{\varepsilon_{\mathbf{k}}\left(\varepsilon_{\mathbf{k}}+2C+2\Delta\right)},\\
E_{\mathbf{k}+} & = & \sqrt{\left(\varepsilon_{\mathbf{k}}+2C\right)\left(\varepsilon_{\mathbf{k}}+2\Delta\right)}.
\end{eqnarray}
$E_{\mathbf{k}-}$ is the gapless phonon spectrum, while $E_{\mathbf{k}+}$
becomes gapped due to the bosonic pairing. As a result of $E_{\mathbf{k}}(v_{\mathbf{k}1}^{2}+v_{\mathbf{k}2}^{2})=(B_{\mathbf{k}}-E_{\mathbf{k}})/2$,
we obtain from Eq. (\ref{eq:OmegaT0}) and Eq. (\ref{eq:OmegaT1})
the thermodynamic potential at finite temperature,
\begin{eqnarray}
\varOmega & = & \varOmega_{0}+\varOmega_{T},\\
\varOmega_{0} & = & -\mathcal{V}\left[\frac{C^{2}}{g}+\frac{\Delta^{2}}{g_{12}}\right]+\sum_{\mathbf{k}}\left[\frac{E_{\mathbf{k}-}+E_{\mathbf{k}+}}{2}-B_{\mathbf{k}}\right],\label{eq:OmegaT0Homo}\\
\varOmega_{T} & = & \frac{1}{\beta}\sum_{\mathbf{k}}\left[\ln\left(1-e^{-\beta E_{\mathbf{k}-}}\right)+\ln\left(1-e^{-\beta E_{\mathbf{k}+}}\right)\right].\label{eq:OmegaT1Homo}
\end{eqnarray}
The quantum and thermal depletions are given by,
\begin{eqnarray}
n_{\textrm{qd}} & = & \frac{1}{\mathcal{V}}\sum_{\mathbf{k}}\left[\frac{B_{\mathbf{k}}}{2E_{\mathbf{k}-}}+\frac{B_{\mathbf{k}}}{2E_{\mathbf{k}+}}-1\right],\label{eq:nqd}\\
n_{\textrm{th}} & = & \frac{1}{\mathcal{V}}\sum_{\mathbf{k}}\left[\frac{B_{\mathbf{k}}}{E_{\mathbf{k}-}}f_{B}\left(E_{\mathbf{k}-}\right)+\frac{B_{\mathbf{k}}}{E_{\mathbf{k}+}}f_{B}\left(E_{\mathbf{k}+}\right)\right],\label{eq:nth}
\end{eqnarray}
respectively.

\begin{figure}[t]
\begin{centering}
\includegraphics[width=0.48\textwidth]{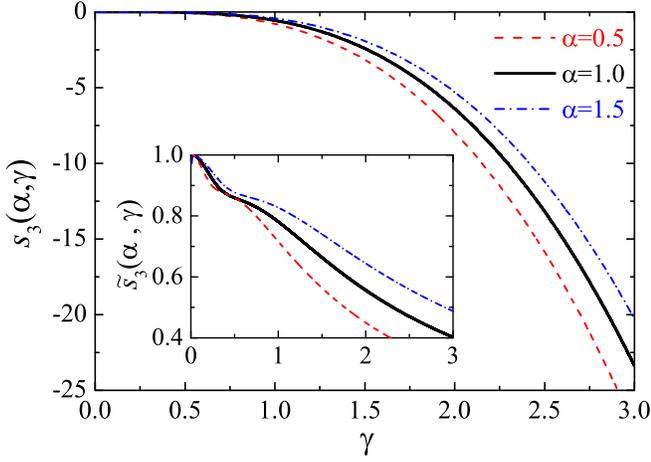}
\par\end{centering}
\caption{\label{fig1_s3} The function $s_{3}\left(\alpha,\gamma\right)$ as
a function of the reduced temperature $\gamma=k_{B}T/C$ at three
different ratios $\alpha=\Delta/C=0.5$ (red dashed line), $1.0$
(black solid line), and $1.5$ (blue dot-dashed line). The inset shows
the function $\tilde{s}_{3}\left(\alpha,\gamma\right)$.}
\end{figure}

\section{Three-dimensional droplets}

In three dimensions, the zero-temperature thermodynamic potential
Eq. (\ref{eq:OmegaT0Homo}) has been discussed in detail in the previous
works \citep{Hu2020a,Hu2020b}. By replacing the bare interaction
strengths $g$ and $g_{12}$ with the $s$-wave scattering lengths
$a$ and $a_{12}$, one obtains \citep{Hu2020a,Hu2020b},
\begin{equation}
\frac{\varOmega_{0}}{\mathcal{V}}=-\frac{m}{4\pi\hbar^{2}}\left[\frac{C^{2}}{a}+\frac{\Delta^{2}}{a_{12}}\right]+\frac{8m^{3/2}C^{5/2}}{15\pi^{2}\hbar^{3}}\mathcal{G}_{3}\left(\frac{\Delta}{C}\right),\label{eq:OmegaT0Homo3D}
\end{equation}
where $\mathcal{G}_{3}(\alpha)\equiv(1+\alpha)^{5/2}+h_{3}(\alpha)$
with $h_{3}(\alpha)\equiv(15/4)\int_{0}^{\infty}dt\sqrt{t}[\sqrt{(t+1)(t+\alpha)}-(t+1/2+\alpha/2)+(1-\alpha)^{2}/(8t)]$
. To carry out the momentum integration in the finite-temperature
contribution to the thermodynamic potential Eq. (\ref{eq:OmegaT1Homo}),
we introduce $t=[\hbar^{2}k^{2}/(2m)]/(2C)$, $\alpha=\Delta/C$ and
$\gamma=k_{B}T/C$ to write the two dispersion relations into the
dimensionless forms,
\begin{eqnarray}
\tilde{E}_{-}(t)=\beta E_{\mathbf{k}-} & = & \frac{2\sqrt{t\left(t+1+\alpha\right)}}{\gamma},\\
\tilde{E}_{+}(t)=\beta E_{\mathbf{k}+} & = & \frac{2\sqrt{\left(t+1\right)\left(t+\alpha\right)}}{\gamma}.
\end{eqnarray}
Hence, we find that,
\begin{equation}
\frac{\varOmega_{T}}{\mathcal{V}}=\frac{8m^{3/2}C^{5/2}}{15\pi^{2}\hbar^{3}}s_{3}\left(\alpha,\gamma\right),\label{eq:OmegaT1Homo3D}
\end{equation}
where 
\begin{equation}
s_{3}\left(\alpha,\gamma\right)\equiv\frac{15}{4}\gamma\int_{0}^{\infty}dt\sqrt{t}\sum_{\pm}\ln\left[1-e^{-\tilde{E}_{\pm}\left(t\right)}\right].
\end{equation}
At very low temperature, i.e., $k_{B}T\ll C$ or $\gamma\ll1$, where
the gapless phonon spectrum can be well-approximated as $\hat{E}_{-}(t)\simeq(2\sqrt{1+\alpha}/\gamma)\sqrt{t}$
and the gapped mode $\tilde{E}_{+}(t)$ does not contribute to the
integral, we obtain the low-temperature result,
\begin{align}
s_{3} & \simeq\frac{15}{4}\gamma\int_{0}^{\infty}dt\sqrt{t}\ln\left[1-\exp\left(-\frac{2\sqrt{1+\alpha}}{\gamma}\sqrt{t}\right)\right],\nonumber \\
 & =-\frac{\pi^{4}}{48}\frac{\gamma^{4}}{\left(1+\alpha\right)^{3/2}}\propto-T^{4}.\label{eq:s3LowTemp}
\end{align}
where in the last step, we have introduced the variable $x=(2\sqrt{1+\alpha}/\gamma)\sqrt{t}$
and have used the identity $\int_{0}^{\infty}dxx^{2}\ln(1-e^{-x})=-\pi^{4}/45$.
Therefore, it is useful to rewrite $s_{3}(\alpha,\gamma)$ into the
form,
\begin{equation}
s_{3}\left(\alpha,\gamma\right)=-\frac{\pi^{4}}{48}\frac{\gamma^{4}}{\left(1+\alpha\right)^{3/2}}\tilde{s}_{3}\left(\alpha,\gamma\right),\label{eq:s3}
\end{equation}
where the function $\tilde{s}_{3}(\alpha,\gamma)$ accounts for the
high-order correction at nonzero temperature.

\begin{figure}[t]
\begin{centering}
\includegraphics[width=0.48\textwidth]{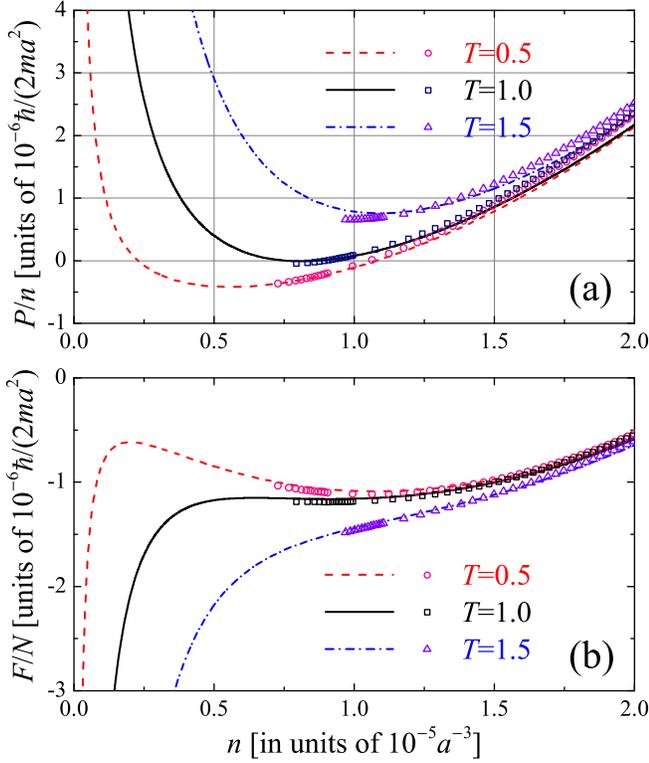}
\par\end{centering}
\caption{\label{fig2_eos3d} Pressure per particle $P/n$ (a) and free energy
per particle $F/N$ (b) of a 3D droplet, in units of $10^{-6}\hbar^{2}/(2ma^{2})$,
as a function of the density $n$ at three temperatures $k_{B}T=0.5$,
$1.0$ and $1.5$ measured in units of $10^{-4}\hbar^{2}/(2ma^{2})$.
The symbols and lines show the numerical and analytical results, respectively.
The density $n$ is in units of $10^{-5}a^{-3}$. We take the inter-species
interaction strength $a_{12}=-1.05a$.}
\end{figure}

In Fig. \ref{fig1_s3}, we show the functions $s_{3}(\alpha,\gamma)$
and $\tilde{s}_{3}(\alpha,\gamma)$ as a function of the reduced temperature
$\gamma$ at three typical ratios $\alpha=\Delta/C=0.5$ (red dashed
line), $1.0$ (black solid line), and $1.5$ (blue dot-dashed line).
We find that $s_{3}(\alpha,\gamma)$ follows closely its low-temperature
approximate result Eq. (\ref{eq:s3LowTemp}) up to $k_{B}T\sim C$,
i.e., when the thermal energy $k_{B}T$ becomes comparable to the
intra-species interaction energy $C$. At this temperature scale,
$s_{3}(\alpha,\gamma)\sim-\mathcal{O}(1)$ is at the same order of
the function $\mathcal{G}_{3}(\alpha)$ but has an opposite sign,
indicating that the repulsive force provided by LHY quantum fluctuations
might be compensated by thermal fluctuations. More quantitatively,
at $\alpha=1$ where $\mathcal{G}_{3}\left(\alpha=1\right)=4\sqrt{2}$,
the combined contribution to the thermodynamic potential from quantum
and thermal fluctuations,
\begin{equation}
\frac{\varOmega_{\textrm{LHY}}}{\mathcal{V}}=\frac{32\sqrt{2}m^{3/2}C^{5/2}}{15\pi^{2}\hbar^{3}}\left[1-\frac{\pi^{4}}{768}\gamma^{4}\tilde{s}_{3}\left(1,\gamma\right)\right],
\end{equation}
vanishes at about $\gamma\simeq2$.

\subsection{Equation of state}

For a given chemical potential $\mu$, we numerically calculate the
thermodynamic potential $\varOmega=\varOmega_{0}+\varOmega_{T}$ as
a function of the pairing gap $\Delta$, by using Eq. (\ref{eq:OmegaT0Homo3D}),
Eq. (\ref{eq:OmegaT1Homo3D}) and Eq. (\ref{eq:s3}). The saddle-point
solution $\Delta=\Delta_{0}$ is then obtained by minimizing the thermodynamic
potential. After the calculation of the number of particles $N=-\partial\varOmega/\partial\mu$,
we finally determine the free energy $F=\varOmega+\mu N$ and the
pressure $P=-\varOmega/\mathcal{V}$.

Quite generally, in three dimensions the saddle-point pairing gap
is much large than the chemical potential $\Delta_{0}\gg\left|\mu\right|$
and hence $C=\mu+\Delta_{0}\gg\left|\mu\right|$ \citep{Hu2020a,Hu2020b}.
Therefore, to a very good approximation, around the saddle point ($\Delta\sim\Delta_{0}$)
we obtain,
\begin{eqnarray}
\frac{\varOmega}{\mathcal{V}} & = & -\frac{m\Delta}{2\pi\hbar^{2}a}\mu-\frac{m}{4\pi\hbar^{2}a}\left(1+\frac{a}{a_{12}}\right)\Delta^{2}\nonumber \\
 &  & +\frac{32\sqrt{2}m^{3/2}\Delta^{5/2}}{15\pi^{2}\hbar^{3}}\left[1-\frac{\pi^{4}}{768}\gamma^{4}\tilde{s}_{3}\left(1,\gamma\right)\right],
\end{eqnarray}
where we have set $C=\Delta$ in the LHY term (i.e., the second line)
and thereby $\gamma=k_{B}T/\Delta$. By taking the derivatives with
respect to the chemical potential $\mu$ and the pairing gap $\Delta$,
i.e., $n=N/\mathcal{V}=-\partial(\varOmega/\mathcal{V})/\partial\mu$
and $\partial\varOmega/\partial\Delta=0$ at $\Delta=\Delta_{0}$,
we find that,
\begin{equation}
\Delta_{0}=\frac{2\pi\hbar^{2}a}{m}n
\end{equation}
and\begin{widetext}
\begin{equation}
\mu=-\left(1+\frac{a}{a_{12}}\right)\Delta_{0}+\frac{32\sqrt{2m}a}{3\pi\hbar}\Delta_{0}^{3/2}\left\{ 1+\frac{\pi^{4}}{1280}\left[\gamma^{4}\tilde{s}_{3}\left(1,\gamma\right)+\frac{2}{3}\gamma^{5}\frac{\partial\tilde{s}{}_{3}\left(1,\gamma\right)}{\partial\gamma}\right]\right\} ,
\end{equation}
respectively. It is then straightforward to obtain the analytic expressions
for the pressure $P$ and the free energy $F$,
\begin{eqnarray}
P & = & -\frac{\pi\hbar^{2}}{m}\left(a+\frac{a^{2}}{a_{12}}\right)n^{2}+\frac{128\sqrt{\pi}}{5}\left(\frac{\hbar^{2}a^{5/2}}{m}\right)n^{5/2}\left\{ 1+\frac{5\pi^{4}}{2304}\left[\gamma^{4}\tilde{s}_{3}\left(1,\gamma\right)+\frac{2}{5}\gamma^{5}\frac{\partial\tilde{s}{}_{3}\left(1,\gamma\right)}{\partial\gamma}\right]\right\} ,\label{eq:Pressure3D}\\
\frac{F}{\mathcal{V}} & = & -\frac{\pi\hbar^{2}}{m}\left(a+\frac{a^{2}}{a_{12}}\right)n^{2}+\frac{256\sqrt{\pi}}{15}\left(\frac{\hbar^{2}a^{5/2}}{m}\right)n^{5/2}\left[1-\frac{\pi^{4}}{768}\gamma^{4}\tilde{s}_{3}\left(1,\gamma\right)\right],\label{eq:FreeEnergy3D}
\end{eqnarray}
\end{widetext}where the parameter $\gamma$ should now be understood
as, 
\begin{equation}
\gamma=\frac{mk_{B}T}{2\pi\hbar^{2}an}=\frac{1}{\left(na^{3}\right)^{1/3}\left[2\zeta(3/2)\right]^{2/3}}\left(\frac{T}{T_{c}}\right).\label{eq:gamma3D}
\end{equation}
In the second step of the above equation, we have re-expressed the
density $n$ in terms of the small gas parameter $na^{3}\ll1$ and
the Bose-Einstein condensation temperature of an ideal gas $k_{B}T_{c}=2\pi\hbar^{2}[n/(2\zeta(3/2))]^{3/2}/m$,
where $\zeta(3/2)\simeq2.6124$ is the Riemann zeta function. In the
zero-temperature limit, the free energy Eq. (\ref{eq:FreeEnergy3D})
recovers the analytical expression for the ground-state energy found
in earlier works \citep{Hu2020a,Hu2020b}. 

In Fig. \ref{fig2_eos3d}, we report the numerical and analytical
results of the pressure per particle $P/n=-\varOmega/N$ (a) and the
free energy per particle $F/N$ (b) as a function of the density $n$,
at the inter-species interaction strength $a_{12}=-1.05a$ and at
three typical temperatures $k_{B}T=0.5$, $1.0$ and $1.5$, measured
in units of $10^{-4}\hbar^{2}/(2ma^{2})$. There is an excellent agreement
between numerical and analytical predictions, as we anticipate. At
finite temperature, we find that the density dependence of the pressure
and free energy develop non-trivial features in the low-density limit
($n\rightarrow0$). Instead of becoming vanishingly small as in the
zero-temperature case, both equations of state become divergent when
the density decreases to zero. This divergence is purely a temperature
effect and can be easily understood from the last term in Eq. (\ref{eq:Pressure3D})
and Eq. (\ref{eq:FreeEnergy3D}), i.e., 
\begin{equation}
\delta P,\delta F\propto n^{5/2}\gamma^{4}=n^{5/2}\left(\frac{mk_{B}T}{2\pi\hbar^{2}an}\right)^{4}\sim\frac{T^{4}}{n^{3/2}}.
\end{equation}
The temperature correction in the pressure and free energy therefore
diverges like $\pm n^{-3/2}$ as the density approaches zero and vanishes
like $T^{4}$ with decreasing temperature.

As a consequence of such a divergent low-density dependence, at low
temperature (i.e., at $k_{B}T=0.5\times10^{-4}\hbar^{2}/(2ma^{2})$
as indicated by the red dashed line) we find there are two solutions
for the self-bound condition $P=0$, which correspond to a local maximum
and a local minimum in the free energy, respectively. However, the
lower density self-bound solution (i.e., the local maximum in the
free energy) is not mechanically stable and should be discarded, since
one can readily identify that its inverse compressibility, 
\begin{equation}
\kappa^{-1}=-\mathcal{V}\left(\frac{\partial P}{\partial\mathcal{V}}\right)_{N}=n\frac{\partial P}{\partial n}<0,
\end{equation}
becomes negative so the system collapses. As we increase temperature,
we observe that the two self-bound solutions start to merge and eventually
disappear at a threshold temperature $k_{B}T_{\textrm{th}}\sim10^{-4}\hbar^{2}/(2ma^{2})$.
Above this threshold temperature, the pressure is always positive
and there is no longer a local minimum in the free energy (see, i.e.,
the blue dot-dashed lines at $k_{B}T=1.5\times10^{-4}\hbar^{2}/(2ma^{2})$).
Thus, the self-bound droplet state is completely destabilized by the
temperature effect.

\begin{figure}[t]
\begin{centering}
\includegraphics[width=0.48\textwidth]{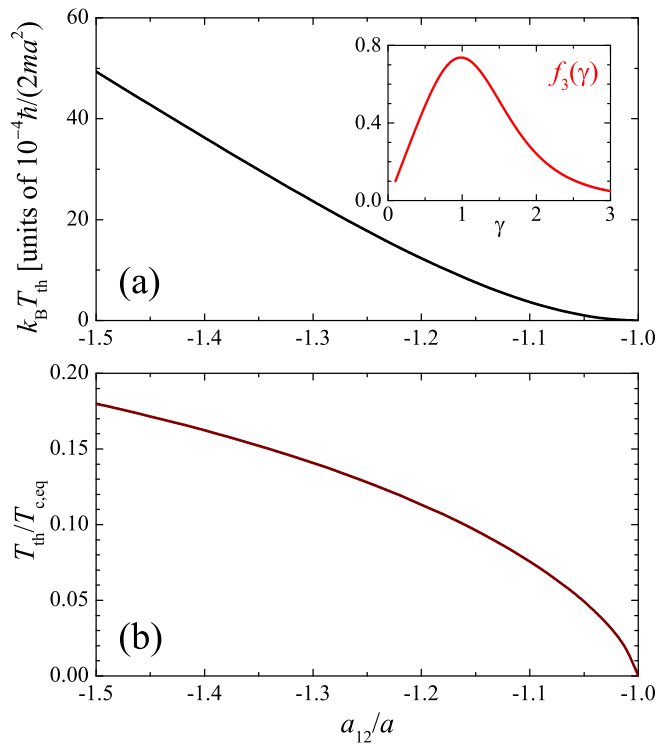}
\par\end{centering}
\caption{\label{fig3_tth3d} Threshold temperature $T_{\textrm{th}}$ of a
3D droplet as a function of the inter-species interaction strength
$a_{12}/a$, measured in units of $10^{-4}\hbar^{2}/(2ma^{2})$ (a)
and in units of the ideal gas BEC transition temperature $T_{c,\textrm{eq}}$
at the zero-temperature equilibrium density $n_{\textrm{eq}}$ (b),
as predicted by the analytic equation (\ref{eq:TTCEQ}). The inset
in (a) show the function $f_{3}(\gamma)$.}
\end{figure}

\begin{figure}[t]
\begin{centering}
\includegraphics[width=0.48\textwidth]{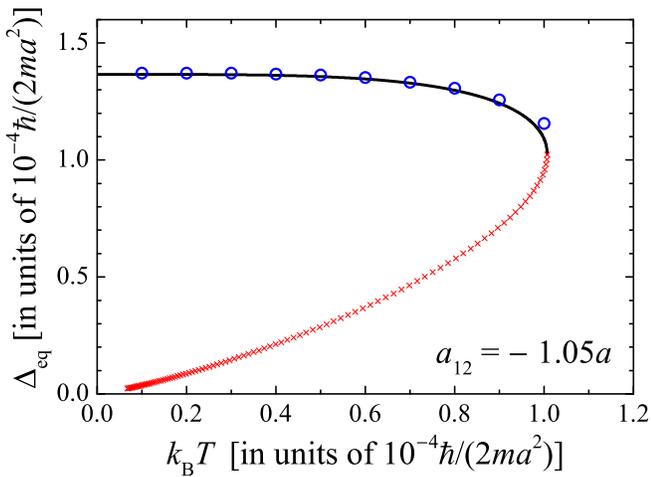}
\par\end{centering}
\caption{\label{fig4_delta} Temperature dependence of the equilibrium pairing
gap $\Delta_{\textrm{eq}}$ of a 3D droplet at the inter-species interaction
strength $a_{12}=-1.05a$. The equilibrium pairing gap $\Delta_{\textrm{eq}}$
and temperature $k_{B}T$ are both measured in units of $10^{-4}\hbar^{2}/(2ma^{2})$.
The blue circles and black line show the numerical and analytical
results, respectively. The red crosses correspond to the unphysical
solution, in which the droplet becomes mechanically unstable with
$\partial P/\partial n<0$.}
\end{figure}

\subsection{Threshold temperature for destabilization}

We may determine the threshold temperature for destabilization from
the analytic expression for the pressure. By setting $P=0$ in Eq.
(\ref{eq:Pressure3D}) and using Eq. (\ref{eq:gamma3D}), we obtain
that
\begin{equation}
k_{B}T=\frac{25\pi^{2}}{8192}\left(1+\frac{a}{a_{12}}\right)^{2}\frac{\hbar^{2}}{ma^{2}}f_{3}\left(\gamma\right),
\end{equation}
where we have defined the function,
\begin{equation}
f_{3}\left(\gamma\right)\equiv\gamma\left\{ 1+\frac{5\pi^{4}}{2304}\left[\gamma^{4}\tilde{s}_{3}\left(1,\gamma\right)+\frac{2}{5}\gamma^{5}\frac{\partial\tilde{s}{}_{3}\left(1,\gamma\right)}{\partial\gamma}\right]\right\} ^{-2}.
\end{equation}
As shown in the inset of Fig. \ref{fig3_tth3d}(a), $f_{3}(\gamma)$
is a non-monotonous function and reaches its maximum 
\begin{equation}
\max_{\gamma}\left[f_{3}(\gamma)\right]\simeq0.7372
\end{equation}
 at $\gamma_{\textrm{th}}\simeq0.9835$. Therefore, we find
\begin{equation}
k_{B}T_{\textrm{th}}\simeq0.0222\left(1+\frac{a}{a_{12}}\right)^{2}\frac{\hbar^{2}}{ma^{2}},\label{eq:TTC3D}
\end{equation}
which is shown in Fig. \ref{fig3_tth3d}(a) as a function of the interaction
strength ratio $a_{12}/a$. By recalling that the zero-temperature
equilibrium density $n_{\textrm{eq}}$ is given by \citep{Hu2020a},
\begin{equation}
n_{\textrm{eq}}=\frac{25\pi}{16384}\left(1+\frac{a}{a_{12}}\right)^{2}a^{-3},
\end{equation}
we can measure the threshold temperature in units of the condensation
temperature $T_{c,\textrm{eq}}$ at the equilibrium density $n_{\textrm{eq}}$,
\begin{equation}
\frac{T_{\textrm{th}}}{T_{c,\textrm{eq}}}\simeq0.3742\left(1+\frac{a}{a_{12}}\right)^{2/3}.\label{eq:TTCEQ}
\end{equation}
As can be seen from Fig. \ref{fig3_tth3d}(b), under the typical experimental
conditions (i.e., $a_{12}\sim[-1.10,-1.05]a$ as in Refs. \citep{Cabrera2018,Semeghini2018}),
the threshold temperature is less than one-tenth of the condensation
temperature. This small threshold temperature can alternatively be
understood from Eq. (\ref{eq:gamma3D}). At the threshold reduced
temperature, $\gamma_{\textrm{th}}\sim1$, we find that $T_{\textrm{th}}/T_{c,\textrm{eq}}\sim(na^{3})^{1/3}\sim0.1$
for the small interaction parameter $na^{3}\sim10^{-5}-10^{-4}$ in
the experiments \citep{Cabrera2018,Semeghini2018}.

In Fig. \ref{fig4_delta}, we present the temperature dependence of
the equilibrium pairing parameter $\Delta_{\textrm{eq}}(T)$ at the
inter-species interaction strength $a_{12}=-1.05a$. As the the pairing
parameter is proportional to the density, this figure also show the
equilibrium density $n_{\textrm{eq}}$ as a function of temperature.
In general, with increasing temperature there are two branches in
$\Delta_{\textrm{eq}}(T)$ and $n_{\textrm{eq}}(T)$. The lower branch,
which is shown by red crosses, corresponds to the unstable low-density
self-bound solution we discussed earlier and therefore should be neglected.
We find that for $T<0.7T_{\textrm{th}}$ the temperature dependence
in the upper branch $n_{\textrm{eq}}$ is relatively weak. However,
towards and upon reaching the threshold temperature, the equilibrium
density $n_{\textrm{eq}}$ can decrease significantly, by several
tens of percent in relative.

\begin{figure}[t]
\begin{centering}
\includegraphics[width=0.48\textwidth]{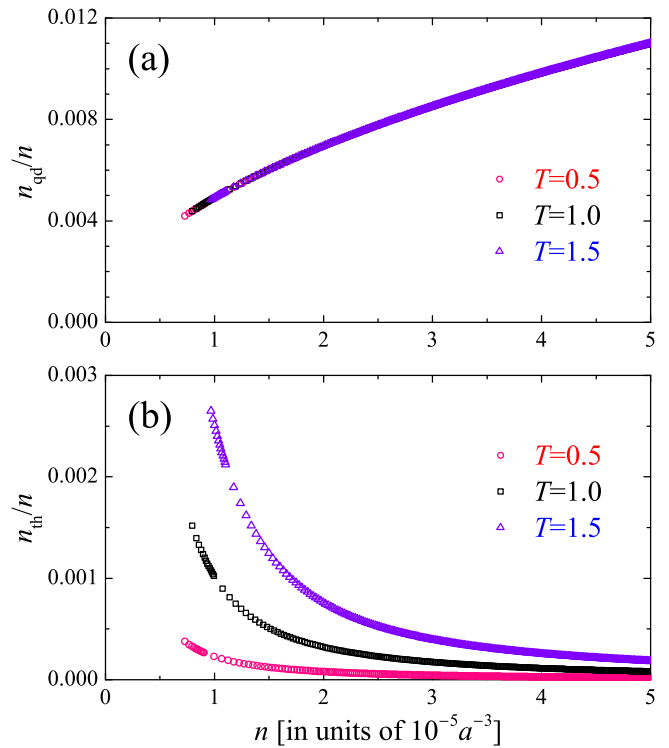}
\par\end{centering}
\caption{\label{fig5_depletion} Quantum depletions (a) and thermal depletions
(b) of a 3D droplet as a function of the density $n$ (in units of
$10^{-5}a^{-3}$) at the inter-species interaction strength $a_{12}=-1.05a$.
We consider three temperatures $k_{B}T=0.5$, $1.0$ and $1.5$, measured
in units of $10^{-4}\hbar^{2}/(2ma^{2})$.}
\end{figure}

\subsection{Quantum and thermal depletions}

The results we presented so far are all obtained within the Bogoliubov
theory, which is valid at sufficiently low temperature in the weakly
interacting regime. To have a self-consistent check, in Fig. \ref{fig5_depletion},
we show the quantum depletion (a) and thermal depletion (b) as a function
of the density at the inter-species interaction strength $a_{12}=-1.05a$,
obtained by using Eq. (\ref{eq:nqd}) and Eq. (\ref{eq:nth}). At
the typical temperatures considered in this figure, the quantum depletion
is less than one percent and is essentially temperature dependent.
The thermal depletion is even smaller and is about $0.1\%$ close
to the threshold temperature for the thermal destabilization of the
droplet state. Therefore, we conclude that the conditions for the
application of the Bogoliubov theory are well satisfied.

\section{One-dimensional droplets}

\begin{figure}[t]
\begin{centering}
\includegraphics[width=0.48\textwidth]{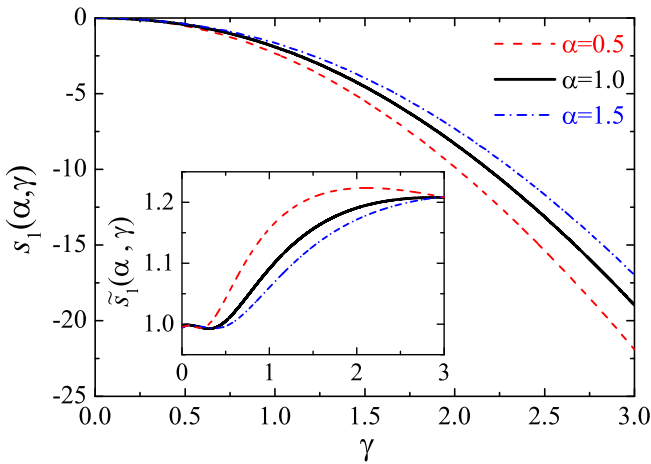}
\par\end{centering}
\caption{\label{fig6_s1} The function $s_{1}\left(\alpha,\gamma\right)$ as
a function of the reduced temperature $\gamma=k_{B}T/C$ at three
different ratios $\alpha=\Delta/C=0.5$ (red dashed line), $1.0$
(black solid line), and $1.5$ (blue dot-dashed line). The inset shows
the function $\tilde{s}_{1}\left(\alpha,\gamma\right)$.}
\end{figure}

We now turn to consider one-dimensional quantum droplets. In this
case, counterintuitively, the droplet formation is driven by LHY quantum
fluctuations \citep{Petrov2016,Parisi2019}, which provide an attractive
force to the system. It is then balanced by the repulsive mean-field
force under the condition $g>-g_{12}$. As given in our earlier work
\citep{Hu2020b}, the zero-temperature thermodynamic potential takes
the form ($\mathcal{V}$ is now the length of the system),

\begin{equation}
\frac{\varOmega_{0}}{\mathcal{V}}=-\left[\frac{C^{2}}{g}+\frac{\Delta^{2}}{g_{12}}\right]-\frac{2\sqrt{m}}{3\pi\hbar}C^{3/2}\mathcal{G}_{1}\left(\frac{\Delta}{C}\right),\label{eq:OmegaT0Homo1D}
\end{equation}
where $\mathcal{G}_{1}(\alpha)\equiv(1+\alpha)^{3/2}+h_{1}(\alpha)$
with $h_{1}\equiv(3/2)\int_{0}^{\infty}dtt^{-1/2}[t+(1+\alpha)/2-\sqrt{(t+1)(t+\alpha)}]$.
As in the three-dimensional case, we introduce the variables $t$,
$\alpha$ and $\gamma$ to rewrite the finite-temperature contribution
to the thermodynamic potential,
\begin{equation}
\frac{\varOmega_{T}}{\mathcal{V}}=\frac{2\sqrt{m}}{3\pi\hbar}C^{3/2}s_{1}\left(\alpha,\gamma\right),\label{eq:OmegaT1Homo1D}
\end{equation}
where 
\begin{equation}
s_{1}\left(\alpha,\gamma\right)\equiv\frac{3}{2}\gamma\int_{0}^{\infty}dt\frac{1}{\sqrt{t}}\sum_{\pm}\ln\left[1-e^{-\tilde{E}_{\pm}\left(t\right)}\right].
\end{equation}
At very low temperature $\gamma\ll1$, where only the gapless phonon
mode contributes to the integral for $s_{1}(\alpha,\gamma)$, we obtain,
\begin{equation}
s_{1}\simeq3\gamma\int_{0}^{\infty}dt\ln\left(1-e^{-2t\sqrt{1+\alpha}/\gamma}\right)=-\frac{\pi^{2}\gamma^{2}}{4\sqrt{1+\alpha}}.\label{eq:s1LowTemp}
\end{equation}
Therefore, it is convenient to rewrite $s_{1}(\alpha,\gamma)$ in
the form,
\begin{equation}
s_{1}\left(\alpha,\gamma\right)=-\frac{\pi^{2}}{4}\frac{\gamma^{2}}{\sqrt{1+\alpha}}\tilde{s}_{1}\left(\alpha,\gamma\right).\label{eq:s1}
\end{equation}
The temperature- or $\gamma$-dependence of $s_{1}(\alpha,\gamma)$
and $\tilde{s}_{1}(\alpha,\gamma)$ at three selected values of $\alpha$
is shown in Fig. \ref{fig6_s1}. In contrast to the three-dimensional
case, we find that the higher-order correction factor $\tilde{s}_{1}$
is generally larger than $1.0$ and does not change too significantly
as we increase the temperature.

By collecting the contributions from quantum and thermal fluctuations
to the thermodynamic potential, we obtain ($\alpha=C/\Delta$ and
$\gamma=k_{B}T/C$),

\begin{equation}
\frac{\varOmega_{\textrm{LHY}}}{\mathcal{V}}=-\frac{2\sqrt{m}}{3\pi\hbar}C^{3/2}\left[\mathcal{G}_{1}\left(\alpha\right)+\frac{\pi^{2}\gamma^{2}}{4\sqrt{1+\alpha}}\tilde{s}_{1}\left(\alpha,\gamma\right)\right].
\end{equation}
It is clear that the contributions from quantum and thermal fluctuations
do not cancel with each other and become comparable at the reduced
temperature $\gamma\sim\mathcal{O}(1)$. Naïvely, one may think that
thermal fluctuations enhance the stability of the the one-dimensional
droplet state, contrary to its three-dimensional counterpart. However,
it turns out to be an incorrect picture.

\begin{figure}[t]
\begin{centering}
\includegraphics[width=0.48\textwidth]{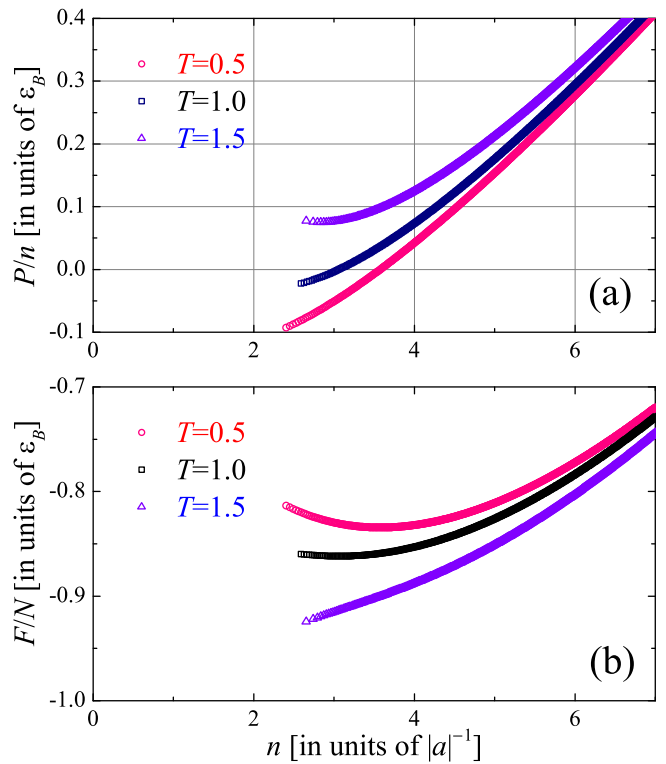}
\par\end{centering}
\caption{\label{fig7_eos1d} Pressure per particle $P/n$ (a) and free energy
per particle $F/N$ (b) of a 1D droplet, in units of $\varepsilon_{B}$,
as a function of the density $n$ (in units of $\left|a\right|^{-1}$)
at three temperatures $k_{B}T/\varepsilon_{B}=0.5$, $1.0$ and $1.5$
and at the inter-species interaction strength $g_{12}=-0.75g$. The
pressure and free energy per particle and temperature are all measured
in units of $\varepsilon_{B}=\hbar^{2}/(ma_{12}^{2})$.}
\end{figure}

\begin{figure}[t]
\begin{centering}
\includegraphics[width=0.48\textwidth]{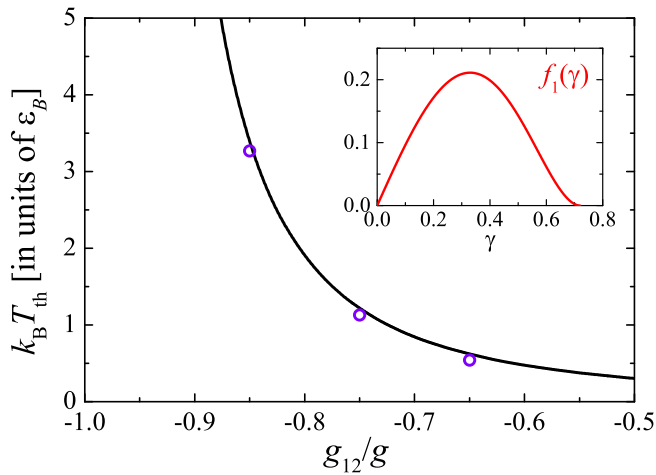}
\par\end{centering}
\caption{\label{fig8_tth1d} Threshold temperature $T_{\textrm{th}}$ (in units
of $\varepsilon_{B}$) of a 1D droplet as a function of the inter-species
interaction strength $g_{12}/g$, predicted by the analytic equation
(\ref{eq:TTC1D}). The empty circles report the results from numerical
calculations. The inset shows the function $f_{1}(\gamma)$.}
\end{figure}

We have performed numerical calculations for the equations of state,
by minimizing $\varOmega=\varOmega_{0}+\varOmega_{T}$ in Eq. (\ref{eq:OmegaT0Homo1D})
and Eq. (\ref{eq:OmegaT1Homo1D}) at a given chemical potential $\mu$
and consequently calculating the pressure and the free energy. Their
numerical results at the inter-species interaction strength $g_{12}=-0.75g$
is reported in Fig. \ref{fig7_eos1d}. As shown in (a) for pressure,
at sufficiently low temperature, i.e., $k_{B}T=0.5\varepsilon_{B}$,
where $\varepsilon_{B}=\hbar^{2}/(ma_{12}^{2})$ is the binding energy
for an inter-species dimer, we find a droplet state satisfying the
self-bound condition $P=0$ (or equivalently a local minimum $F/N$)
at the density $n_{\textrm{eq}}\sim3\left|a\right|^{-1}$. With increasing
temperature ($k_{B}T=\varepsilon_{B}$), the equilibrium density $n_{\textrm{eq}}$
becomes smaller. At the largest temperature considered in the figure,
i.e., $k_{B}T=1.5\varepsilon_{B}$, it seems that the pressure $P$
is always positive and at the same time the free energy per particle
decreases monotonically as the density decreases to zero. Therefore,
the self-bound droplet state disappears at a threshold temperature
$\varepsilon_{B}<k_{B}T_{\textrm{th}}<1.5\varepsilon_{B}$.

To understand the thermal destabilization of the droplet state, it
is useful to derive an analytic expression for the pressure, following
the same steps in the three-dimensional case. However, we note that,
the condition $\left|\mu\right|\ll C,\Delta_{0}$ is not so well-satisfied
in one dimension \citep{Hu2020b}. As a result, the analytic expressions
derived are of qualitative use only (so we do not show them in Fig.
\ref{fig7_eos1d}). For the pressure $P$, it is straightforward to
obtain that,
\begin{eqnarray}
P & = & -\frac{1}{4}\left(g+\frac{g^{2}}{g_{12}}\right)n^{2}-\frac{\sqrt{m}g^{3/2}}{3\pi\hbar}n^{3/2}+\nonumber \\
 &  & \frac{\pi\sqrt{m}g^{3/2}}{16\hbar}n^{3/2}\left[\gamma^{2}\tilde{s}_{1}\left(1,\gamma\right)+\frac{2}{3}\gamma^{3}\frac{\partial\tilde{s}_{1}\left(1,\gamma\right)}{\partial\gamma}\right],\label{eq:Pressure1D}
\end{eqnarray}
where the last two terms come from quantum fluctuations and thermal
fluctuations, respectively. It is easy to see that, while the contribution
from quantum fluctuations to the pressure is negative, the contribution
from thermal fluctuations is always positive and increases with increasing
temperature, consistent with our numerical results shown in Fig. \ref{fig7_eos1d}(a).
Therefore, thermal fluctuations eventually destroy the droplet state.

By setting $P=0$ in Eq. (\ref{eq:Pressure1D}), we find the relation,
\begin{equation}
k_{B}T=\frac{32}{9\pi^{2}}\frac{\varepsilon_{B}}{\left(1+g_{12}/g\right)^{2}}f_{1}\left(\gamma\right),
\end{equation}
where
\begin{equation}
f_{1}\left(\gamma\right)\equiv\gamma\left\{ 1-\frac{3\pi^{2}}{16}\left[\gamma^{2}\tilde{s}_{1}\left(1,\gamma\right)+\frac{2}{3}\gamma^{3}\frac{\partial\tilde{s}{}_{1}\left(1,\gamma\right)}{\partial\gamma}\right]\right\} ^{2}.
\end{equation}
The function $f_{1}(\gamma)$ is shown in the inset of Fig. \ref{fig8_tth1d}.
There is a maximum 
\begin{equation}
\max_{\gamma}\left[f_{1}(\gamma)\right]\simeq0.21116
\end{equation}
 at $\gamma_{\textrm{th}}\simeq0.330$. Therefore, the threshold temperature
in one dimension is given by,
\begin{equation}
k_{B}T_{\textrm{th}}\simeq0.0761\frac{\varepsilon_{B}}{\left(1+g_{12}/g\right)^{2}}.\label{eq:TTC1D}
\end{equation}
This analytical prediction is shown in Fig. \ref{fig8_tth1d} by using
a black solid line. We have also numerically determined the threshold
temperature by repeating the calculations in Fig. \ref{fig7_eos1d}
at different inter-species interaction strengths and show the results
in the figure by empty circles. It seems that there is a good agreement
between numerical and analytical results. In the interval of interest,
i.e., $g_{12}\sim[-0.9,-0.5]g$, the threshold temperature is roughly
at the order of the binding energy $\varepsilon_{B}$ of the inter-species
dimer.

\section{Conclusions}

In summary, we have presented a systematic investigation of the temperature
effect in self-bound ultradilute quantum droplets, both in three dimensions
and in one dimension, by extending a recently developed microscopic
pairing theory \citep{Hu2020a,Hu2020b} to nonzero temperature. We
have shown that thermal fluctuations generally destabilize the droplet
state and destroy it above a threshold temperature. The energy scale
of the threshold temperature is at the order of the intra-species
interaction energy and is therefore very small in the weakly interacting
regime. For a three-dimensional quantum droplet, the threshold temperature
is less than one-tenth of the Bose-Einstein condensation temperature
under current experimental conditions \citep{Cabrera2018,Semeghini2018}.
We have also predicted the temperature dependence of the equilibrium
density and have found that it can decrease by several tens of percent
upon reaching the threshold temperature. 

Our predictions could be readily examined in future experiments on
quantum droplets realized by a binary Bose mixture with attractive
inter-species interactions. The sensitive temperature dependence of
the droplet state may alternatively provide us good thermometry in
ultracold atomic experiments, where the low temperature at the scale
of one-tenth of the condensation temperature is often challenging
to measure.
\begin{acknowledgments}
This research was supported by the Australian Research Council's (ARC)
Discovery Program, Grants No. DE180100592 and No. DP190100815 (J.W.),
Grant No. DP170104008 (H.H.), and Grant No. DP180102018 (X.-J.L).
\end{acknowledgments}

\end{document}